# Next generation clinical trials: Seamless designs and master protocols

Abigail Burdon[a,*], Thomas Jaki[a,b], Xijin Chen[a], Pavel Mozgunov[a], Haiyan Zheng[a,c], Richard Baird[d]

a MRC Biostatistics Unit, University of Cambridge, UK
b University of Regensburg, Germany
c Department of Mathematical Sciences, University of Bath, UK
d Department of Oncology, University of Cambridge, UK

*Correspondence to: MRC Biostatistics Unit, University of Cambridge, Cambridge CB2 0SR, United Kingdom.
E-mail address: abigail.burdon@mrc-bsu.cam.ac.uk (A. Burdon)
ORCID identification: 0000-0002-0883-4160

## Abstract

**Background:** Drug development is often inefficient, costly and lengthy, yet it is essential for evaluating the safety and efficacy of new interventions. Compared with other disease areas, this is particularly true for Phase II / III cancer clinical trials where high attrition rates and reduced regulatory approvals are being seen. In response to these challenges, seamless clinical trials and master protocols have emerged to streamline the drug development process.
**Methods:** Seamless clinical trials, characterized by their ability to transition seamlessly from one phase to another, can lead to accelerating the development of promising therapies while Master protocols provide a framework for investigating multiple treatment options and patient subgroups within a single trial.
**Results:** We discuss the advantages of these methods through real trial examples and the principals that lead to their success while also acknowledging the associated regulatory considerations and challenges.
**Conclusion:** Seamless designs and Master protocols have the potential to improve confirmatory clinical trials. In the disease area of cancer, this ultimately means that patients can receive life-saving treatments sooner.

# Background

Development of novel medicines to treat disease is a long and costly process. It is also risky, with most drugs never achieving regulatory approval. This white paper focuses on the area of oncology, in which attrition rates are particularly high [1]. Overall success rates are roughly 14% across all disease areas compared with 1% in oncology trials alone [1]. In recent years, an increased emphasis has been placed on data-informed decision-making during the development of oncology treatments; in the meanwhile, a more quantitative assessment of risk is used when making decisions about subsequent investments in studies [2].

With a seamless clinical trial design, different stages of development (e.g., Phase I / II, or Phase II / III) are combined in a single trial protocol. This can streamline the study not only operationally but inferentially (i.e., to include data from Phase II in the decision at the end of Phase III) [3]. It can also potentially speed up the development, while allowing decisions about particular treatment(s) as promising or not promising ones to be made quickly. Some drugs in rare disease areas, such as cancer, may qualify for accelerated approval in which decisions are based on surrogate endpoints such as biomarkers. This kind of regulatory approval may allow for early approval and similarly speed up the drug development process.

Another class of clinical trial design which can improve the efficiency of clinical drug development are master protocols, in which multiple treatments, and cancer subtypes, can be tested within a single trial protocol.

In this manuscript we discuss different modern approaches that have the potential to make decisions about a treatment's value more efficiently. As a consequence, this might help deliver life-saving oncology treatments to patients more quickly.

# Phase II/III seamless trials

In clinical drug development, the initial testing of new drugs in Phase II trials helps identify which treatments are looking most promising - and potentially worth testing in (expensive, long) confirmatory Phase III trials.

Conventionally, the learning and confirmatory phases are separate in both their planning and data analyses. One way to accelerate the development process is to combine these phases into a single protocol, so the between-phase 'white space' can be minimised or even removed. Such 'seamless' designs require careful planning in advance, including decision rules and patient recruitment strategies, etc. Nonetheless, seamless designs are recognised as more efficient in risk management. For instance, interim decision gates can be incorporated to terminate the evaluation early, if the treatment is either not sufficiently effective, or associated with unacceptable side effects.

The past decades have witnessed a steady increase in the number of seamless trials. Some prominent examples include the AGILE trial [4] treating patients with glioblastoma, and the

Horizon III trial [5] in the disease area of advanced colorectal cancer. Among seamless designs, there are two distinct types: *operationally seamless* and *inferentially seamless*. In an operationally seamless trial, the data collected from the learning and confirmatory stages are kept separate. Here, benefits mainly arise from the combination of the two phases into a single protocol and the reduction in time for decision making. Inferentially seamless trials allow for learning phase data to contribute to the analysis at the confirmatory stage and help establish efficacy. Such trials are particularly efficient due to the re-use of data, which can potentially lead to reduced sample sizes. Inferentially seamless trials are considered to be a subset of operationally seamless trials (Figure 1).

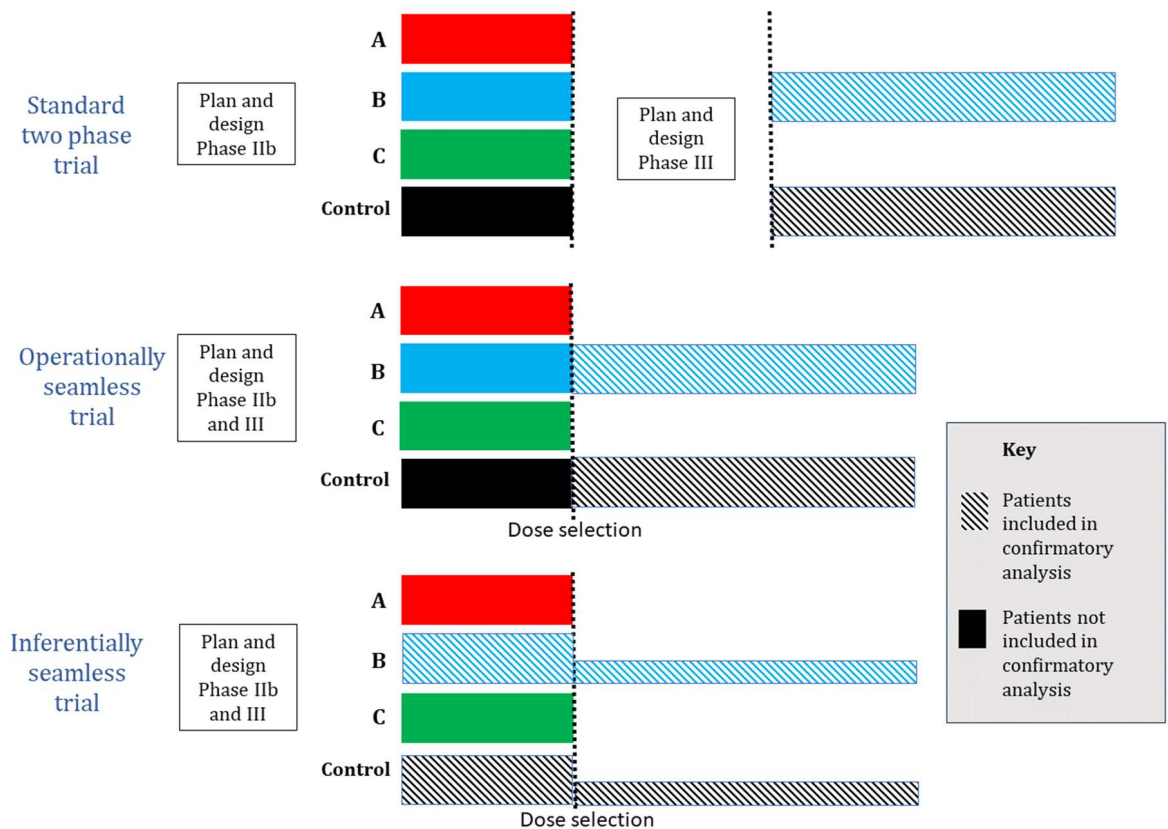

Figure 1: Illustration of inferentially seamless and operationally seamless Phase IIb / III trials.

Seamless designs can facilitate the investigation of a few clinically important questions in parallel within a single study. In Phase II of a clinical development program, investigators may be interested in comparing multiple dose levels or schedules, with the aim being to identify the most efficacious one(s). It is then of interest to compare the experimental treatment, at a chosen dose, with the current standard of care option in a confirmatory Phase III study [6]. Alternatively, Phase II may have been designed with the focus being subgroup identification [7]. In such trials, pre-defined subgroups of patients are given both experimental and control treatments and the trial would aim to identify which subgroup(s) respond well to treatment. For example, in metastatic breast cancer studies, it is well recognised that treatments specifically targeting HER2 hormone receptors are particularly effective [8]. A seamless Phase II / III study, instead, allows for two questions to be answered succinctly and efficiently without pausing to assess the results of Phase II on completion. In an inferentially seamless trial, the data obtained on patients who received the selected dose or were in the benefitting subgroup(s) can be used in the analysis before and after the adaptation. Seamless trials also allow for added efficiency by exploiting information from short-term endpoints in Phase II to aid the inference for, and at, the confirmatory phase.

Consequently, the inferentially seamless trial must be set up in a way to respond quickly to interim decisions. For example, the Phase III recruitment strategy should be planned prior to Phase II, for all possible combinations of treatment and subgroups, and this is unlikely to be identical across different subgroups. Further, a well designed clinical trial builds in flexibility to respond to changes. This can be incorporated by critically evaluating and anticipating possible complexities by computer simulations performed at the outset. The risk of unplanned change is even greater for seamless studies. Thus, the need for increased upfront investment and planning is apparent.

For example, in order to control the type 1 error rates in inferentially seamless trials, there should be careful consideration regarding hypothesis testing at each phase. The null hypothesis here is that the experimental treatment is equivalent to control in terms of efficacy. When data is collected and the hypothesis is tested after each phase, multiple testing procedures should be employed to ensure that type 1 error rates are not inflated overall [9]. It may be desirable for more weight to be given to the results of the Phase III study in comparison to Phase II, or for concluding results based on Phase II only to require proportionally high levels of evidence. These scenarios can be achieved by using a higher significance level in Phase II than in Phase III which is known as "alpha splitting". There are many options for the choice of alpha splitting which can be tailored towards the investigator's priorities and popular choices include the Pocock [9], O'Brien and Fleming [10] and Lan de Mets boundaries [11]. This poses an additional design consideration for seamless designs compared with classical approaches.

During the analysis of the data from each phase, recommendations can be made to the sponsor regarding how the trial should proceed in the form of go/no-go decisions. However, to construct these rules, it is necessary for at least one person to have access to data with labelled treatment groups. This unblinding can lead to biased results and compromise the integrity of the drug development. The role of an independent data monitoring committee (IDMC) [12] is therefore imperative to ensure that seamless trials are rigorously conducted. The IDMC may review unblinded data to assess the go/no-go criteria. Another crucial role of the IDMC is to monitor safety which may require unblinding data at the patient level. For traditional non-seamless trials, there is no consequence for unblinding data at the end of Phase II since the analysis refreshes and type 1 error rates are effectively re-set. Hence, the compromise of unblinding individuals should be taken into account at the design phase of a seamless clinical trial.

Standard practice in oncology clinical trials is to focus on some long-term endpoint, such as overall survival (OS) or progression free survival (PFS), as the primary outcome of interest at the end of Phase III. When collecting observations on these long-term endpoints, the timing of the analyses must be carefully considered. One must ensure that sufficient information is available, while being ethical about the length of time that patients are receiving a potentially inefficacious treatment. In many trials, rapidly available short-term endpoints may be collected and analysed to conduct treatment and/or subgroup selection at the end of Phase II. Data that contribute to these different endpoints can be combined to create a seamless Phase II / III trial which efficiently informs decisions across phases [13]. Hence, we do not need to restrict seamless designs to trials where the Phase II outcome is the same long-term endpoint as in Phase III. However, using data from multiple endpoints can result in complex analysis methods, thus requiring increased planning. For example, trialists are warned of introducing bias and inflating type 1 error rates when the endpoints are correlated and one solution is to separate the data into non-overlapping time periods for each endpoint [14]. This work is naturally extended to incorporate Phase IV, time-to-event endpoints in master protocol studies [15]. Although the statistical challenges are heightened, there is much to be gained when we allow endpoints to differ across phases of seamless trials.

## Example: Horizon III

The Horizon III trial [5] was an inferentially seamless Phase II / III study which compared a new treatment, cediranib with mFOLFOX6, against the standard of care, bevacizumab with mFOLFOX6, in patients with advanced colorectal cancer. During Phase II, patients were randomised among three arms to receive 20mg of cedirinib, 30mg of cedirinib or a placebo alternative. At the end of Phase II analysis, the primary outcome, overall response rate (ORR) was reported and the IDMC concluded that the treatment arm of 20mg of cedirinib met predefined safety and efficacy criteria. At the start of Phase III, patients who were previously enrolled in Phase II on either 20mg of cedirinib or placebo continued treatment and newly enrolled patients were randomised to either of these two treatment arms. Furthermore, patients who were previously receiving 30mg of cedirinib were unblinded and given the opportunity to be re-randomised into the trial. At the end of Phase III analysis, the primary outcome was PFS, which did not meet the predefined inferiority limit. Hence, it was concluded that cediranib activity was comparable to that of bevacizumab. Importantly, there were 225 patients whose data contributed towards decisions in both Phases II and III and it was estimated that the drug was made available roughly 1–2 years sooner [5] than a conventional trial which does not incorporate the seamless aspect.

# Master protocols

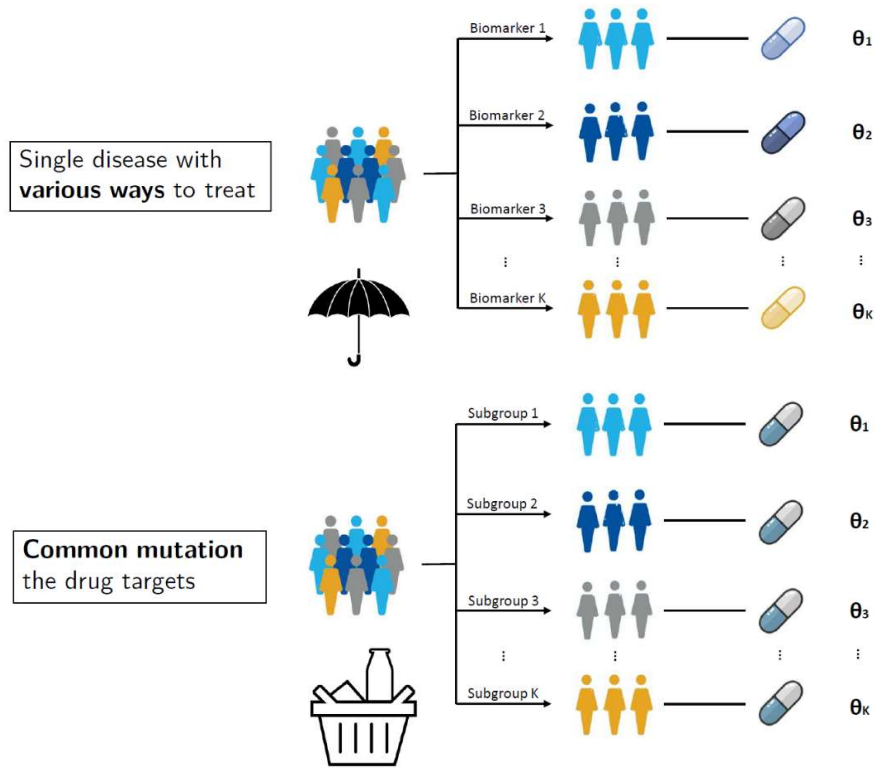

Figure 2: Illustration of two types of master protocols, i.e. basket trials and umbrella trials.

The most frequently used clinical trial design is the *two-parallel-group design*, in which participants are randomised to one of two treatment groups and then followed over time for assessment of outcomes. There are situations when the evaluation of multiple experimental treatments in parallel is desired. One approach would be to do a series of two-parallel-group trials, each one comparing one experimental treatment to the control. On the other hand, efficiencies can be anticipated by evaluating the multiple treatments in a *multi-arm design*. Arms refer to treatments that include the investigational drug at one or more doses, and one or more control treatments such as placebo and/or an active comparator. In this design, patients are assigned to one of several experimental treatments or controls and will stay in their assigned treatment arm for the duration of the study.

Taking this one step further, we come to master protocols that allow for multiple research questions to be answered under one unifying trial protocol. The benefits include the assessment of multiple treatments and/or patient subgroups, enabling an ongoing trial to add or drop treatments and subgroups. Generally, these designs go beyond the multi-arm trial due to their time-effective features to further expedite the clinical drug development process. Table 1 briefly describes the master protocol designs that we shall discuss in this paper.

**Table 1: Definitions of some types of master protocol designs within oncology trials.**

| Type of Master protocol design | Definition |
|---|---|
| **Platform** | Multiple experimental treatments assessed in parallel and further treatments can be added and/or removed after the study has started. |
| **Basket** | Tests the effectiveness of a new treatment in patients of various cancer types, sharing a common mutation or biomarker, to which the new treatment targets. |
| **Umbrella** | Patients are enrolled of the same cancer type but different gene mutations. The trial evaluates the effectiveness of multiple treatments that are linked with these mutations and biomarkers. |

For each of the master protocol designs, the null and alternative hypotheses are more complicated than both the two-parallel-group designs and multi-arm designs, as there are a number of possibilities to consider [16]. The considerations for these trials require additional attention as the number of treatment arms, subgroups and hypotheses are increased. Efficiency gains can also be made by combining types of master protocols, for example we can have Basket type trials sequentially joining an overarching platform and also Umbrella and Basket methodologies can be combined into a single trial protocol.

## Platform trials

Platform trials are one of the new approaches to clinical research which address the need for enhanced operational efficiency in the modern era of increasingly specific cancer subpopulations and decreased resources to study treatments for individual cancer subtypes in a traditional way [17]. A platform trial comprises a platform in which multiple treatments or treatment combinations are evaluated against a common control for a single disease in homogeneous patient populations in a perpetual manner. The focus of platform trials is on the disease, including several disease subtypes, rather than any particular experimental therapy. Investigated treatments are allowed to enter or leave the platform, often according to pre-specified decision rules.

Platform trials have been successfully implemented or are currently being planned in a variety of diseases, including breast cancer [18,19], lung cancer [20-22] and brain cancer including both Phase II and Phase III settings [17,23]. There are some prominent platform trials such as the Phase II I-SPY 2 [24,25] and the multi-arm multi-stage (MAMS), STAMPEDE used in prostate cancer [26,27].

Within platform trials, different treatments can be dropped for futility and be added as they become available. Thus, allowing for exploring combinations of treatments and directly comparing competing treatments, both of which are commonly ignored in premarket settings. On the other hand, traditional trial designs are proposed by a single sponsor to evaluate a prespecified and generally limited number of treatments in a homogeneous group of patients. The sharing of resources in platform trials among different sponsors can greatly reduce costs and increase statistical efficiency. This type of trial design has demonstrated great potential for the need for enhanced operational efficiency across the broad spectrum clinical research. Firstly, not all experimental treatments need be included in the randomization scheme at the start of the trial minimising starting delays. Recruitment speed can be improved compared to starting a new trial from scratch as patients have an increased chance of being allocated an experimental treatment. Platform trials also have the ability to adapt to time trends, for example allowing the standard of care to be updated with accumulating evidence. Finally, smaller sample sizes are required due to a shared control group.

Despite these benefits that have been discussed widely in the literature, corresponding challenges have attracted attention during the implementation of platform trials. Challenges include regulatory acceptance [28], issues concerning the trial management [29], data management [30] and statistical properties such as error rate control, mid-trial shifts in the randomization ratios, and sample size calculation, etc. [31].

## Basket trials

Basket trials involve testing a single new treatment across multiple diseases / cancer types in a single clinical trial protocol. The application in oncology involves enrolling patients with a certain molecular feature (e.g., DNA mutation), irrespective of the location or origin of cancer. Figure 2 gives a graphical illustration of basket trials. One representative example is a Phase II basket trial (NCT01524978) evaluating the efficacy of vemurafenib in patients with BRAF V600-mutant, a targetable oncogene in various nonmelanoma cancers [32]. This vemurafenib basket trial comprises seven sub-studies defined by distinct cancer histologies, with the results suggesting preliminary vemurafenib activity in some, but not all, nonmelanoma cancers.

Considering such patient heterogeneity, analysis of basket trials is often directed towards the reporting of the sub-study specific treatment effects. Although stand-alone analyses fully acknowledge the heterogeneity, regarding the sub-studies in isolation means low-powered tests due to their small sample sizes. Many sophisticated statistical methods have been proposed to enable borrowing of information [33-35], as those could be justified by the biological basis that all sub-studies share a commonality (e.g., BRAF V600-mutant in the vemurafenib trial). Major advantages of such approaches generally include a higher statistical power to detect early efficacy of the treatment, and a smaller sample size [36] than is required by standard-alone analyses to ensure the same level of decision accuracy.

Basket trials are proven to possess enhanced efficiency in Phase II settings, for the simultaneous evaluation of multiple patient subgroups. Further efficiency may be gained by incorporating interim monitoring, as subgroups that are less likely to respond can be dropped at the mid-course. More complex configuration of basket trials that accommodate multiple targets and/or agent combinations have also been performed (e.g., see Basket of Basket trials below).

## Umbrella trials

Umbrella trials are another major type of precision cancer medicine trials [37], in which patients with a single tissue type of cancer (e.g., breast cancer) have molecular profiling performed on their tumours to test for the presence of multiple different biomarkers. Accordingly, umbrella trials involve a few individualised treatments to see if they might each be matched to the biomarker. Patients with tumours positive for biomarker A are matched to drug A, those positive for biomarker B to drug B, and so on.

The molecular profiling performed in many umbrella trials involves the DNA sequencing of tumours to test for potentially targetable gene mutations which may be driving the growth of those cancers. A pioneering example of such studies is the SAFIR-01 trial [38], in which 423 patients with metastatic breast cancer consented to have a fresh tumour biopsy for profiling by comparative genomic hybridisation and Sanger gene sequencing. 195 of these patients were found to have targetable genomic alterations and 55 were started on molecular targeted therapy matched to their tumour.

Umbrella trials can be highly efficient since they permit the parallel testing of multiple therapies in a single clinical trial protocol. One of the major challenges of umbrella trials is that many targetable gene mutations are uncommon, and may be found in only 1-5% of patient cancers. Therefore, a huge logistical effort is required to screen enough patients to find those with these alterations which can be successfully targeted.

## Example: I-SPY 2 trial

Investigation of serial studies to predict your therapeutic response with imaging and molecular analysis 2 (I-SPY 2) is a Phase II trial used to screen neoadjuvant treatment of women with locally advanced breast cancer [39]. Multiple therapies are tested as neoadjuvant treatments in I-SPY 2. Unlike 'treatment-focused' designs, I-SPY 2 is a 'disease-focused' master protocol trial [40], which is in the context of several sub-studies for different disease subtypes [41]. Response-adaptive randomization is used to assign patients to the most promising treatment in their respective genetic breast-cancer subgroups while maintaining a sufficient number of patients assigned to the standard of care [37]. For this reason, this biomarker-based Bayesian adaptive design achieves the improved efficiency in identifying improved treatment regimens for patient subsets on the basis of molecular characteristics (biomarker signatures) [39]. In the trial, regimens will be dropped for reasons of futility if they show a low probability of improved efficacy with any biomarker signature. New drugs will enter as those that have undergone testing are graduated or dropped.

## Example: Basket of Baskets trial

The Basket of Baskets trial ('BoB', NCT) is an example of a Master Protocol in which patients first undergo broad DNA-sequencing of their cancer ('BoB iProfiler'). Patients with certain targetable DNA abnormalities can then be offered a therapy matched to their tumour profile. There are multiple different targeted therapies available within the trial protocol - each effectively its' own basket trial - hence 'Basket of Baskets'. Treatments include an immune checkpoint inhibitor module for patients with tumours bearing genetic features thought to increase the chance of immunotherapy response (including high Tumour Mutational Burden and individual mutations in genes involved in DNA damage repair). New treatment arms are added to the BoB trial protocol as protocol amendments, making efficient use of the existing trial protocol and molecular profiling infrastructure.

## Regulatory Approvals

Full regulatory approval for drugs to treat cancer requires solid evidence of safety and therapeutic effectiveness based on adequate and well-controlled clinical trials. Regular approval was the only pathway supported by the US Food and Drug Administration (FDA) until 1992. At that time (in the middle of the HIV crisis), the FDA introduced accelerated approval (AA) as another approval pathway.

Accelerated approval is an accelerated pathway to regulatory approval for drugs that treat serious or life-threatening conditions, including cancer. In order for clinical trial data to qualify for AA, drugs must show an effect on a (surrogate) endpoint which is reasonably likely to predict significant clinical benefit - compared to existing available therapies.

The potential for accelerated approval or conditional marketing authorization based on promising Phase II data [42] has also reshaped how the evaluation of drugs is planned. The route via accelerated approval is not without risk, however. In their review Chen et al. [43] found that only 55% of accelerated approvals based on response rate were converted to full approvals, highlighting the need for fit for high quality data and decision frameworks.

Since there is some risk that the apparent benefit measured by these surrogate endpoints might not be real, drugs given AA must all be further studied in post-approval clinical trials to confirm meaningful clinical benefit. The FDA have published their experience over the first 25 years of accelerated approval, and found that only a small proportion of indications under the AA program fail to verify clinical benefit [44]. For the other drugs where AA benefits have been confirmed, these drugs have been brought to patients years before confirmatory trials would typically have been completed.

Similar systems for expedited regulatory review exist in other parts of the world (see Table 2). In Europe, the European Medicines Agency (EMA) conditional marketing authorisation (introduced in 2006) can be based on less comprehensive evidence at the time of initial authorisation.

**Table 2: Characteristics of US Food and Drug Administration (US FDA) and European Medicines Agency (EMA) expedited programs\**

| Expedited program | Year introduced | Criteria for new drug | Benefits |
|---|---|---|---|
| **FDA** | | | |
| Fast track | 1987 | Potential to address unmet medical need | Earlier interaction with FDA |
| Accelerated approval | 1992 | Advantage over available therapies shown on surrogate endpoint reasonably likely to correlate with meaningful clinical benefit | Quicker approval based on surrogate endpoint |
| Priority review | 1992 | Potential significant improvement in safety or effectiveness | Quicker FDA review time (6 months vs. standard 10 months) |
| Breakthrough therapy | 2012 | Substantial improvement on clinically significant endpoint over available therapies | Intensive support from FDA on rapid drug development |
| **EMA** | | | |
| Accelerated assessment | 2005 | Innovative drug with potential for significant impact on public health | Quicker EMA review time (150 days vs. standard 210 days) |
| Conditional marketing authorisation | 2006 | Potential benefit to public health of rapid availability outweighs risk of less comprehensive clinical trial data | Quicker approval based on surrogate endpoint |

(Table modified from [45])

# Discussion

Although a vast number of options of innovative designs are being used in practice and getting the attention, there is no "all-size-fits-all" approach when it comes to which trial design (master or seamless or both) should be used. The best-suited answer depends on the objective and on a particular trial setting in question. For example, basket designs originally were designed to deal with the problem of small sample size in each subgroup and to enhance the efficiency via borrowing on information across subgroups. However, there should be sufficient scientific rationale and preliminary evidence on why "borrowing" of information might be appropriate between different cancer types (e.g., due to a common mutation). Similarly, although platform trials do provide a great potential for the gain in efficiency, they also impose major inferential challenges (e.g., robustness to time trends, use of non-concurrent controls, etc.). To make the decision on why the platform trial is appropriate, one should weigh these against the option of conducting a "conventional" trial, which might be more robust to these.

Concerning the choice of the seamless design, the decision whether to choose it should match the overall objective of the trial development. For example, if there is a major uncertainty between Phase II and Phase III in the choice of endpoints, populations, doses, durations, etc., it might be beneficial to conduct two separate trials. If different endpoints are used for Phase II and Phase III trials, one should have a solid understanding of the association between these to answer the relevant research question. As for any adaptive design, the evaluation window for the endpoints and recruitment rates would have a major impact on the benefits such seamless options can provide.

Importantly, neither master protocol nor seamless design should be seen as an option to "get away" with no pre-planned analyses and "following the data". Regardless of the trial design used, the decision-making criteria (e.g., to progress to Phase III from Phase II, or to terminate a basket, or to stop trial earlier for efficacy) should be pre-specified before the trial start.

Finally, the consideration on opting to master protocol and/or seamless designs could be driven by the fact whether the study is industrially or academically funded as this majorly affects the ultimate objective. If planning for these, under the academic funding scheme specifically, one needs to ensure the sufficient funding available to reach a definite conclusion on the research question.

# Acknowledgements

For the purpose of open access, the author has applied a Creative Commons Attribution (CC BY) licence to any Author Accepted Manuscript version arising. The views expressed in this publication are those of the authors and not necessarily those of the NHS, the National Institute for Health and Care Research for the Department of Health and Social Care (DHSC).

# Author contributions

# Declaration of Competing Interest

The authors declare that they have no known competing financial interests or personal relationships that could have appeared to influence the work reported in this paper.

# Funding

This project has received funding from the European Union's Horizon 2020 research and innovation programme under grant agreement No. 965397. HZ's and RB's contribution to this manuscript was supported by Cancer Research UK (RCCPDF\100008). This report is independent research supported by the National Institute for Health and Care Research (NIHR300576). PM and TJ also received funding from the UK Medical Research Council (MC_UU_00002/14). Infrastructure support is acknowledged from the NIHR Cambridge Biomedical Research Centre (BRC-1215-20014) and Cambridge Experimental Cancer Medicine Centre.